# Lock-in & Break-out from Technological Trajectories:
# Modeling and policy implications

Wilfred Dolfsma[i] & Loet Leydesdorff[ii]

**Abstract.** Arthur [1, 2] provided a model to explain the circumstances that lead to technological lock-in into a specific trajectory. We contribute substantially to this area of research by investigating the circumstances under which technological development may break-out of a trajectory. We argue that for this to happen, a third selection mechanism— beyond those of the market and of technology—needs to upset the lock-in. We model the interaction, or mutual shaping among three selection mechanisms, and thus this paper also allows for a better understanding of when a technology will lock-in into a trajectory, when a technology may break-out of a lock-in, and when competing technologies may co-exist in a balance. As a system is conceptualized to gain a (third) degree of freedom, the possibility of bifurcation is introduced into the model. The equations, in which interactions between competition and selection mechanisms can be modeled, allow one to specify conditions for lock-in, competitive balance, and break-out.

[i] University of Groningen School of Economics and Business, PO Box 800, 9700 AV Groningen, The Netherlands, w.a.dolfsma@rug.nl

[ii] Amsterdam School of Communications Research (ASCoR), Kloveniersburgwal 49, 1012 CX Amsterdam, The Netherlands, http://www.leydesdorff.net ; loet@leydesdorff.net

**Introduction**

While there has been debate about whether innovations are integrated into systems nationally, regionally, or sectorially focusing on specific technologies such as biotechnology [3, 4], the systemic character of innovation and technological development itself is hardly contested since Dosi [5]. Unlike the neo-classical economics approach of comparative statics, with a focus on market forces creating equilibria at each moment in time, evolutionary economists have emphasized the historical character of innovation processes [6, 7, 8, 9]. According to Schumpeter [10] and Nelson & Winter [8, 9], market equilibria are disturbed by innovations made possible by technological developments over time. The recurrence of this disturbance term may lead to rigidities in the market along trajectories [5, 7]. Evolutionary economics has, however, acknowledged the role of market demand influencing how a technological paradigm emerges and develops as well [11, 12].

Selection by the market and stabilization along a technological trajectory over time may lead to the development of a technological paradigm that can be locked-in. Lock-in of technologies into a paradigm has been argued to be potentially sub-optimal [13, 14, cf. 15]. For a lock-in to emerge and transform (stabilize) into a 'natural trajectory' or 'technological regime' other relevant selection mechanisms need to have reinforced the selection [9: pp. 258f.]. Negative reinforcement, however, is also possible and empirically observable: not all technologies are permanently locked-in into a trajectory [16, 17, 18, 19].

Lock-ins between two subdynamics in a process of mutual shaping can discourage or even prevent new developments. For example, when the market is reinforced by network externalities among previous adopters, the development of a new generation of technologies can be irrelevant to the techno-economic system which prevails. A third selection mechanism impinging on decision-making about relevant combinations at the interfaces between supply-side opportunities and demand-side expectations [20] may then upset a locked-in trajectory. In the case of three relevant selection environments interacting a break-out from a lock-in can also be modeled.[1] In the formal model developed in this paper, we elaborate substantially on the work of Arthur and others.[2]

Like markets and technologies, markets and political decision-making processes can co-evolve along trajectories. Analogously, technologies and political decision-making processes can also be locked-in into state apparatuses. The three different selection mechanisms are equivalent *a priori*.[3] Which technology will be locked-in emerges from the interaction among the three subsystems in potentially unstable conditions, possibly independently of the inherent qualities of the competing technologies.[4] There need thus not be a decision by a specific actor in the system for lock-in or break-out to occur, but rather a non-linear dynamics at the systems level can generate these effects [29].

The paper proceeds as follows. Section 1 sets out the classical Arthur framework as an important backdrop for our formal modeling and policy recommendations. Especially Arthur's [2] implicit assumption, not given sufficient recognition, that lock-in results from two selection environments moving in lockstep is further developed in this paper. Section 2 shows how interaction with a third selection environment can be modeled. Section 3 then derives conditions for break-out from a lock-in that Section 4 further elaborates into options for policy makers in governments and strategy makers in firms.

**1. Arthur's Model of Lock-in**

Brian Arthur [1, 2] specified the mechanism for lock-in in the case of two competing technologies with randomly arriving adopters following their given preferences in a market with marginally increasing returns. Each adopter changes the situation for adopters thereafter, as it can be more attractive to buy a brand or type of product that is common in the market-place than one that is rare. Such a market in this sense is self-organized. The recursive mechanism in the aggregation of "network externalities" leads *necessarily*, according to Arthur, to a lock-in in the long run. Arthur used the example of the VCR: if one particular standard (e.g., VHS rather than Betamax) is increasingly accepted, one passes a point of "no return" and soon video-stores will only have tapes on their shelves for the dominant type of video recorder only.[5]

---

[1] Our argument does not depend on assumptions about the characteristics of a technology, nor on the extent to which agents' knowledge is perfect or complete.
[2] Methodologically, a formal model provides an advance over numerical simulation [as in 21].
[3] We thus develop a model which is driven by the possibility of increasing returns due to network effects. Thus, like Arthur and others who study lock-in of technologies, we do not include the possibility of agents responding to their (social) environment because they (1) are unclear about their pay-off structure [cf 22, 23], (2) because of a need to share knowledge and information [cf. 24], or (3) because of a need to maintain a position [25].
[4] As decision-making involves the factor time, modeling decision-making as the third selection mechanism, allows one to understand technology and product life-cycles as an outcome of co-evolutionary dynamics [10, 26, 27, 28]. Due to limitations of space, this line of thought cannot be pursued.

In Arthur's (1988) formal model two types of agents, R and S, with different "natural inclinations" towards two competing technologies, *A* and *B*, are present. In Table 1, $a_R$ represents the natural preference of R-type agents towards technology *A*, and $b_R$ their (in this case, lower) inclination towards *B*: $a_R > b_R$. Analogously, one can attribute parameters $a_S$ and $b_S$ to S-type agents ($b_S > a_S$). The network effects of adoption (*r* and *s*) are modeled as coefficients to the number of previous adopters of the respective technologies ($n_A$ and $n_B$).

|  | Technology *A* | Technology *B* |
|---|---|---|
| R-Agent | $a_R + rn_A$ | $b_R + rn_B$ |
| S-Agent | $a_S + sn_A$ | $b_S + sn_B$ |

**Table 1:** Returns for agents R and S to adopting technology *A* or *B*, given $n_A$ and $n_B$ previous adopters of *A* and *B* [2: 118]

The values of the cells in Table 1 indicate the return that an agent receives for adoption of the respective technology. In addition to the satisfaction of obtaining the technology of one's choice—that is, following a natural inclination—the attraction of a technology is increased by previous adopters with a term *r* for each R-type agent, and *s* for S-type agents. If R-type and S-type agents arrive on the market randomly, the theory of random walks predicts that competition will eventually lead to a lock-in on either side (*A* or *B*) due to these positive feedback loops.

---

[5] Of course, the 'thick description' of this case provides intricacies that may not all be modeled [28].

For example, agent S would prefer to buy technology *B* (since $b_S > a_S$), but when the total return for buying technology *A* ($a_S + sn_A$) is larger than the return for buying technology *B* ($b_S + sn_B$) this agent will nevertheless buy technology *A*. In formula format, the condition for this lock-in is:

$$a_S + sn_A > b_S + sn_B \qquad (1)$$

The condition for the lock-in into the other technology can be specified, *mutatis mutandis*.

The model is elegant and can easily be programmed as a computer simulation. Using simulations, it can be shown that lock-ins are robust against changes in parameter values by orders of magnitude [29]. Once in place, a lock-in will not normally lead to a break-out to the other technology, or to a return to a competitive balance between the two technologies. Strong reduction of the network effect for the winning technology (in our case, *r*) or, alternatively, enhancement of the network effect of the losing technology (in this case, *s*), even by orders of magnitude, is not likely to change a locked-in configuration.

Simulation results presented in Figure 1 show that if one forces a break-out by further increasing (or decreasing) parameter values by orders of magnitude, the replacement pattern reverts to the curve for lock-in of the *other* technology. In Figure 1, substitution was forced when technology *A* was the previously locked-in technology. These rapid, but ordered substitution processes have, for example, been noted by [30] in seventeen cases of technological substitution. Their finding was that the rate of substitution in all the cases, once begun, did not change throughout its history, but was to be considered as a systems property. Our simulations confirm their findings.

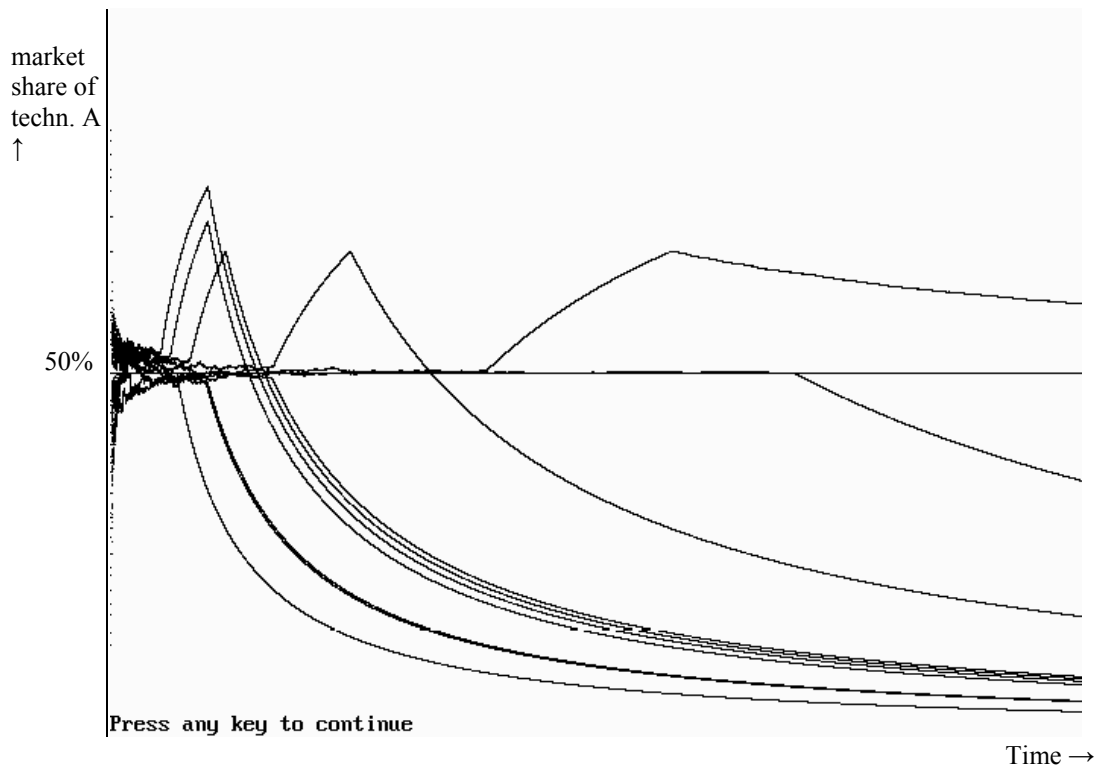

**Figure 1:** Forcing technological break-out from a lock-in and possible return to equilibrium (20,000 adopters) [29, p.316].

In summary, the rate of adoption of a specific technology is not proportional to historical advancements in the technology, but is determined by an evolutionary mechanism underlying substitution. The dynamics are a result of interaction effects among dimensions at the systems level. The dissolution of a lock-in is, in other words, not determined by the emergence of a new and superior technology, but by the *balance* between the interlocking networks of markets and the attractiveness of the technologies for users [31]. Co-evolving sub-systems may thus suppress related sub-dynamics [32].

As Figure 1 indicates (rightmost curve), there is the possibility of a return to a competitive balance between technologies, but only if the market is sufficiently large. Changes in the parameter values can sometimes cause a return to balance or a cascading all the way to a lock-in into the other technology. We would like, however, to suggest that there is a wider set of circumstances that allow for a break-out from a lock-in, and that the process of break-out is in need of a better conceptual understanding.

## 2. Three Selection Environments

The development of a single system $x$ in an environment can be modeled using the logistic equation as follows:

$$x_{t+1} = ax_t(1 - x_t) \quad ; \quad 0 < x < 1 \quad (2)$$

In biology, this equation has been used to model the growth of a population. The feedback term $(1 - x_t)$ inhibits the further growth of this system as the value of $x_t$ increases over time.[6] In the case of a techno-economic system, the technological development generates variation on the basis of the previous state of the system (modeled above as $ax_t$), but this variation is selected by a market environment.[7]

For populations or technologies which compete, one can generalize the logistic equation to the so-called Lotka-Volterra equation in which competition coefficients ($\alpha$) are added to the selection mechanisms [37; 38; 39; 40, p.98]. In both the logistic equation and the Lotka-Volterra equation, selection is represented by a feedback term. This feedback is modeled as $k - \alpha x$ in the case of Lotka-Volterra. Assuming unity of the parameters and without loss of generality it can be modeled as $(1 - x)$. Two selections operating on a technological variation $v$ (defined above as $ax_t$) therefore result in the following quadratic expression for the resulting selection environment:

$$\begin{aligned} f(x) &= v(1-x)(1-x) \\ &= v(x^2 - 2x + 1) \end{aligned} \quad (3)$$

This selection environment no longer operates homogenously, and can be represented as a system with two selections that stabilize at the minimum of the quadratic curve (Figure 2a, left-hand side). When this minimum is extended along the time dimension, a valley is shaped in which the system develops along a trajectory [41, 42]. Emerging rigidities then increasingly structure a system [43].

---

[6] This so-called 'saturation factor' generates the bending of the sigmoid growth curves of systems for relatively small values of the parameter ($1 < a < 3$). For larger values of $a$, the model bifurcates (at $a >= 3.0$) or increasingly generates chaos ($3.57 < a < 4$).

[7] Leydesdorff [35] provided an extension of this model for anticipatory systems—that is, systems which select with hindsight [36] —but this elaboration would unnecessarily complicate the issues here.

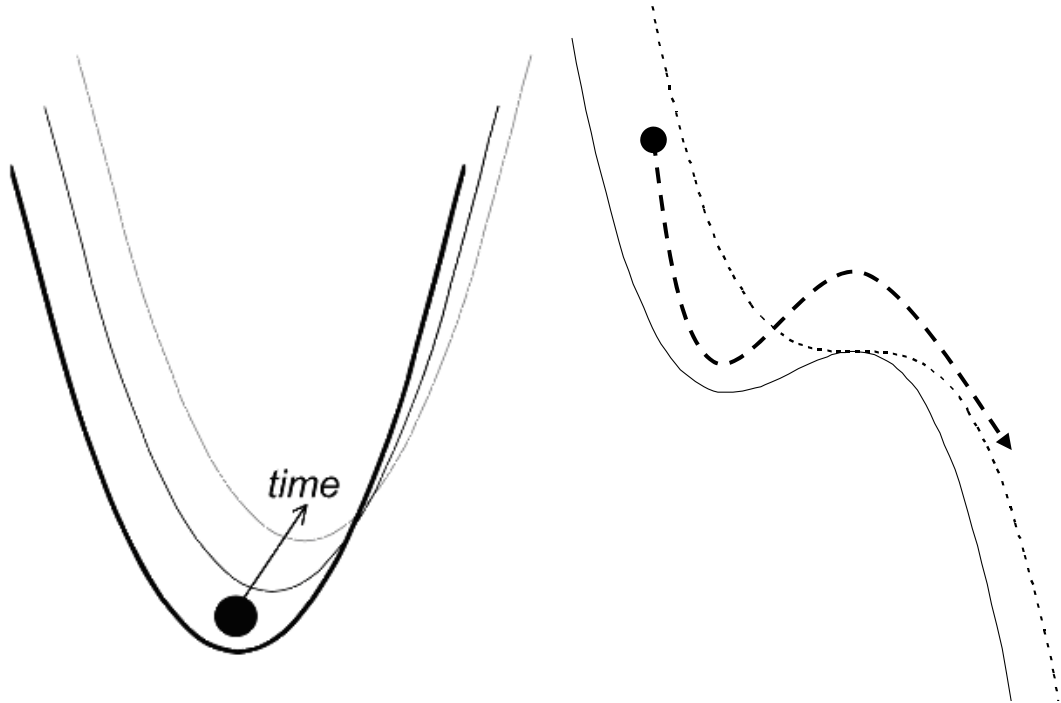

**Figure 2a**:
——— $f(x) = v (x^2 - 2x + 1)$ (stabilized)

**Figure 2b**:
- - - - - $f(x) = -v (x^3 - 3x^2 + 3x - 1)$
——— $f(x) = -v (x^3 - cx^2 + dx - e)$

One may add another selection term to the equation, leading, analogously, to:

$$f(x) = v (1-x)(1-x)(1-x)$$
$$= -v (x^3 - 3x^2 + 3x - 1) \qquad (4)$$

This is represented in Figure 2b. As long as the different selection mechanisms operate synchronously, with the same parameters, a saddle point will emerge (dotted line in Figure 2b). Such a system contains a single (natural) trajectory or regime. The drawn line in Figure 2b shows the configuration resulting when the different selection mechanisms operate with dissimilar parameter values. In this (more general) case the cubic curve shows both a maximum and a minimum.

At the minimum, the techno-economic system is stabilized, but at the maximum a bifurcation can be induced: the system can go left to the relatively stable (local) minimum. If the system goes to the right, the regime will cascade along another trajectory, locking-in into a technological standard more globally. A system may remain stable at the minimum, but the possibility of information entering the system and upsetting the relatively stable (local) configuration may tend to move the system towards a global lock-in. One can also express this as the transition from a local trajectory to a global regime [5]. The local optimum can be considered a niche in which the technology may develop momentum to reach the maximum whereupon it may lock-in at the level of the global market [44, 18]. (We will model this in the next section as a bifurcation.) A niche where two selection enviroments co-evolve may be more or less robust to an upset from a third environment.

The sign of the equations merits attention. Equation 3 had a positive sign, and consequently the hyperbola in Figure 2a showed a minimum. If another sub-dynamics comes into play, this sign is reversed (Equation 4): this inversion cannot be endogenous to only *two* co-evolving selection mechanisms which stabilized the system. A third selection mechanism needs to come into play, a mechanism that either can reinforce the prevailing equilibrium (Figure 3b), or that can invert the sign and make the system susceptible to bifurcation (Figure 3a).

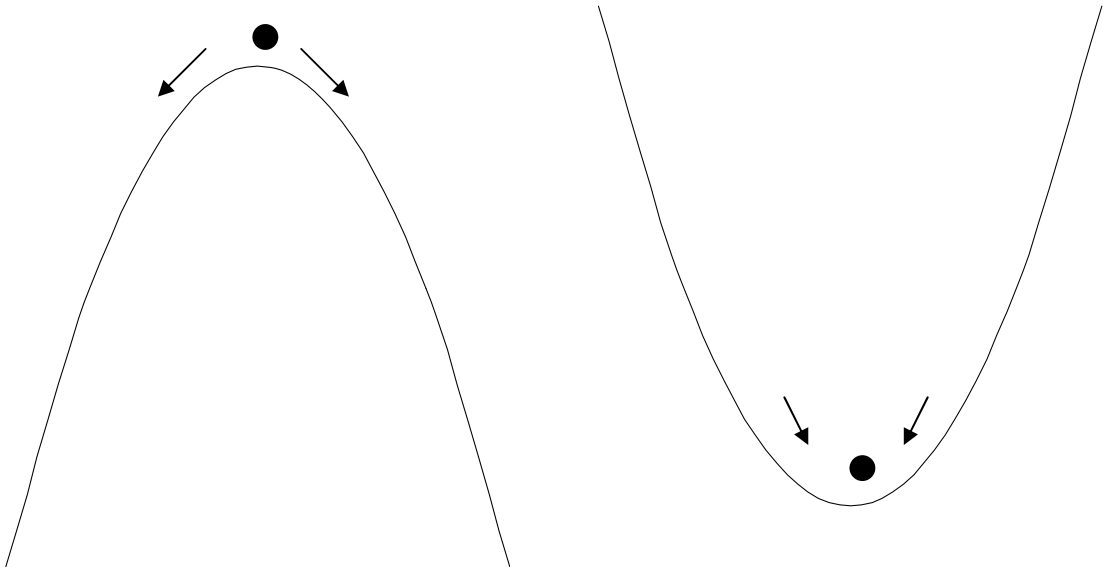

**Figure 3**: Bifurcation (Figure 3a) and hyper-stabilization or lock-in (Figure 3b) of the techno-economic systems in complex selection environments

Figure 3a can be appreciated as a market with increasing marginal returns forcing a system in one of two directions, as, e.g., in Arthur's [1] case of information and communication technologies. Increasing marginal returns leads to a bifurcation as the system at the vertex of the hyperbole is vulnerable to the smallest of effects and will be led to a lock-in on either side. A newly emerging situation is thereafter hyper-stable due to the co-evolution between two sub-dynamics: a technology is locked-in, as in Figure 3b.

**3. Lock-in and Break-out**

Sub-optimal conditions due to a technology that is locked-in into a sub-optimal trajectory is a well-known, though debated, phenomenon in technological development [13, 14, 15]. The QWERTY keyboard, for example, was invented in order to prevent jamming of the keys in the case of mechanical typewriting. For various reasons (e.g., network externalities and learning curves; [6]), the subsequent lock-in could not be reversed after the conditions for this technological fix of the jamming problem had disappeared.

Although the use of the QWERTY keyboard in the case of an electronic computer is sub-optimal, attempts to reverse the prevailing lock-in have failed hitherto.[8]

Studies and models of lock-in have focused on the market (consumption) side or on the technology (production) side, rather than on the interaction between these two sub-dynamics. Under the neo-classical economic assumptions of free competition and open markets, these two sub-dynamics are distinguished as analytically independent, that is, as market clearing and technological innovation. Shifts along the production function due to changing factor prices are distinguished from changes of the production function towards the origin due to technological progress [10, 46].

Unlike the neo-classical framework, the perspective of evolutionary economics initially explicitly addressed the two sub-dynamics of trajectory formation and selection environments as the subjects of theorizing [8, 9]. The feedback terms were recognized by Nelson & Winter, but provisionally black-boxed [8, p. 49]. In their model, trajectory formation—specific organization over time—is induced by the interaction of relevant environments and resources in entrepreneurial practices (e.g., routines). Entrepreneurial practices were considered as the "natural" operators of the economic system [47]. Alternative yet largely compatible explanations have focused on how trajectories form in terms of dominant designs [e.g., 28, 48, 49, 50]. The mechanism of lock-in provides a formal model of the emergence of such a dominant design.

Break-out from a lock-in is possible, in practice as well as in theory, as noted earlier in this article. Arthur [1, 2] showed that potential adopters of a certain technology take the number of previous adopters of competing technologies into account when choosing, even though they may have particular (given) preferences for a technologies. Lock-in can then not be avoided once an adoption threshold is passed. Unlocking of a lock-in can be considered in a more comprehensive, formal manner, however, rather than in an ad hoc one. An additional, third context that becomes relevant as another selection environment may interact with the two co-evolving environments. The interaction among three such selection environments can be modeled formally.

---

[8] David's thesis of an irrational 'lock-in' of a dominant technology has been opposed by [16], who defend the notion of market equilibrium as a basic premise of economic theorizing. Learning curves can be steep, however [45], and competition under increasing returns tends to amplify small historical events that favor one technological option over another [14].

In the case of three interacting sources of variance a complex regime can be expected to emerge. Recombination of three sub-dynamics may generate various types of chaotic behavior [51, 52, 53]. Complex dynamics of three sources of variation can generate lock-ins between two of the sources only if the third context is stable and compatible with the others. This observation is consistent with Kauffman's NK-model [54, 55]. Dynamically, however, the third selection mechanism may erode previously existing rigidities [56]. Return to a competitive balance may be triggered by a historical event such as the advent of radically new technologies, possibly from a different technological or market context [57]. Such historical events are, of course, exogenous to evolutionary modeling. Modeling of a third selection environment implies, however, that return to competitive balance can also be endogenized into the (evolutionary) model [58].

The problem of how a tight coupling between dynamics can be dissolved can be modeled as reaction-diffusion dynamics which have been elaborated in the natural sciences [59, 60, 61, pp. 182ff.].[9] If two systems are tightly coupled (as in a co-evolution; see Figure 5), a simple coupling mechanism can, for example, be specified by the following differential equations:

$$dx_1/dt = -ax_1 + D(x_1 - x_2) + S \qquad (5a)$$
$$dx_2/dt = -ax_2 + D(x_2 - x_1) + S \qquad (5b)$$

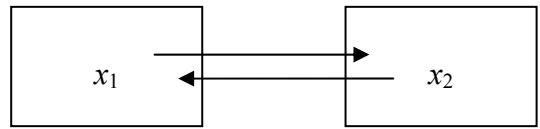

**Figure 5:** Two coupled processes [61, p.183].

---

[9] These insights have hitherto not pervasively influenced the context of economics or other social sciences [34].

Let us assume that $x$ is produced in both compartments at a constant and equal rate $S$. The parameter $a$ represents the decay of $x$; $D$ is the diffusion constant across the interface. (For the sake of simplicity, these parameters are assumed to be equal on both sides.) The diffusion is asymmetrical depending on the concentrations of $x_1$ and $x_2$ in the two compartments. This system of equations provides values for the steady state at:

$$x_1^* = x_2^* = S/a \qquad (6)$$

The concentrations of $x$ in the two environments are then equal, the system is homogeneous. The operational stability of the system, however, is determined in general by the eigenvalues of the matrix of the coefficients of $x_1$ and $x_2$ in Equations 5a and 5b. This matrix is:

$$\begin{Vmatrix} D-a & -D \\ -D & D-a \end{Vmatrix}$$

The two eigenvalues of this system are:

$$\lambda_1 = -a; \qquad \lambda_2 = 2D - a \qquad (7)$$

While the first eigenvalue is always negative ($\lambda_1 = -a$), the second can become positive if $D > a/2$. Thus, if diffusion of $x$ to the other system becomes more important than the flux in the production process (divided by two), a positive and a negative eigenvalue coexist. The system then becomes unstable because a saddle point is generated in the phase diagram (as in Figure 2b). Any deviation from homogeneity will then be amplified, and the system can go through a phase transition, changing the dynamics of the system irreversibly; for example, locking-into a single particular technology.

In the case of two previously coupled dynamics, the bifurcation leads to a polarization, that is, a situation in which all the materials are either in the one cell or the other. Which sub-dynamic will prevail depends on the initial (and potentially random) deviation from homogeneity, possibly provided by the third environment. This more abstract mechanism explains both lock-in and break-out. Lock-in into either Technology *A* or Technology *B* as discussed in the previous section, can be triggered by a random event in a system which has become meta-stable (see Figure 3a above). Break-out can be caused by the increasing diffusion dynamics within a firmly coupled, and therefore locked-in, system. As the diffusion parameter D increases, the previously existing co-evolution may become endogenously unstable [cf. 62]. The trajectory along which the locked-in system develops drives it to increasing diffusion of the technology, and therefore, paradoxically, in so doing, generates the erosion of the very conditions of the lock-in. The next-order globalizing system functions as an attractor [63].

## 4. Conditions for Break-out From Lock-in

To understand the possibility of a return to a competitive balance, consider the analytical conditions for the lock-in specified in Eq. 1. Lock-in into technology *A*, for example, occurs when it becomes more attractive for *S*-type agents to buy this technology despite their natural preference for technology *B*. From Table 1 we derived Eq. 1 which specifies that a lock-in is possible if [2: 120f.]:

$$a_S + sn_A > b_S + sn_B \qquad (1)$$

Thus:

$$sn_A - sn_B > b_S - a_S$$
$$(n_A - n_B) > (b_S - a_S)/s \qquad (8)$$

As we will develop below, from this follow suggestions, first, for preventing a lock-in, and, second, for a break-out from a lock-in, as a third selection environment comes into play disturbing the co-evolution of the two interlocked selection environments.

This elaboration of the model provides certain key insights that can inform both government policy[10] and strategic decision making by firms. Given values for the parameters of natural inclinations and network effects, lock-in is a consequence of the difference in the number of previous adopters for two competing technologies. This difference is ($n_A$ - $n_B$) if technology *A* is the leading technology, and ($n_B$ - $n_A$) in the opposite case. With increasing diffusion, the difference |$n_A$ - $n_B$| becomes smaller as a percentage of the total number of adopters ($n_A$ + $n_B$). In the case of large market shares for both technologies, the difference in market share is more difficult to assess for newly arriving consumers, and they will be more likely to decide on the basis of their natural preferences and thus to prolong competitive balance between the technologies.

In other words, lock-in is easier to prevent in larger markets, and, paradoxically, is more likely to occur in the early stages of new technological developments. If a player such as a government wants to maintain a competitive balance between rivaling technologies, it will need to step in strategically early on in the game, for example, by ostensibly buying the technology that may appear to be loosing. There are, of course, other ways in which governments can enlarge markets for a technology. Opening up markets geographically or allowing technologies to be used for more and different purposes are among them.

Firms as strategic agents have an interest in preventing lock-in, forcing a break-out, or ensuring competitive balance as well. Our model can thus also be considered as a contribution to analyzing competitive relations between firms. Examples of this are airline companies buying airplanes from both Airbus and Boeing, or local and national governments buying the Linux operating system to counter the dominance of the Windows OS in the private sector. Neither of these buying parties would like one supplier to be able to monopolize the market based on the technical characteristics of the products it offers and may be willing to accept technically inferior or more expensive goods from alternative suppliers to prevent lock-in as well as what economists call hold-up [65]. Alternatively, an actor such as a government may prevent knowledge about the relative market shares from becoming available to adopters, to the extent possible. The way in which information about a market becomes available can have profound effects on their functioning [66]. By keeping alternative technologies in the race until markets are well established, lock-in can be postponed or even prevented [67]. Since competitors can be expected to jump on the bandwagon of a technology that gets locked-in, even to the extent of destroying previous competences [68], this may prevent undue disappearance of competing technologies and the competences to further develop them for future use possibly in a different context [18].

---

[10] Our argument, though compatible with that of Metcalfe [64], approaches the issues from a different angle and thus allows us to suggest different considerations for policy.

This happened, for example, to the V2000 and the Betamax video systems competing with the VHS system sponsored by JVC and Matsushita. The products offered by Philips and Sony, respectively, while technically superior, were not accepted in the market as the locked-in dominant design. Philips and Sony both abandoned this area of technical expertise altogether. The competition between different standards for video recording is an appealing case that is much referred to, but it is also a case with a number of additional angles to it [30]. VHS became the dominant technology during the early 1980s and well into 1990s, even though technically it was not superior. The CD did not change this, as video material could not be recorded on it. The DVD, however, became increasingly relevant as an alternative, but this did not mean that the lock-in was immediately broken. A prevailing system can be resilient. After a while, however, when the DVD-share grew independently for other reasons (e.g., due to its superior data storage characteristic), the system tilted and a substitution process generated an cascading away from the VHS towards an entirely new system dynamics. In terms of our modeling, this should be seen as a grown market, affecting the value of $D$ in Equation 5 above. The newly emerging lock-in can be expected to follow the curve of the alternative technology (Figure 1). In terms of the visualization of Figure 2b, the system moves over the hilltop and cascades into a new basin of attraction.

When a lock-in has occurred, the only other option may sometimes be a radical innovation affecting the structural (selection) conditions for the technology. The relevant contexts of the technology need to be changed in order to induce a break-out. Changes at the level of a system can also occur when a new technology that has more pervasive product characteristics becomes available. DVDs, for example, can be considered as a systems innovation when compared with VCRs [41] as they can be used for both video recording and data storage to serve a market much larger than the previous one [69]. Radical innovation implies that one breaks out of one's trajectory, 'creatively destroying' in-house competencies and network externalities.

The right-hand side of Equation 8 suggests that, if the network parameter $s$ of the *losing* technology (e.g., technology *B* as the preference of *S*-type agents) is reduced to zero, the locked-in system is necessarily "unlocked" and reverts to a competitive balance between the rivaling technologies. When $s = 0$, *S*-type agents are free from the constraints of a previous lock-in, and are thereby enabled to return to their natural preference ($b_S$) for Technology B. In Figure 4 the parameter $s$ is set to zero whenever technology *A* captures more than two-thirds (66.7%) of the market. Under these conditions, the system always returns to competitive balance. This discussion is reflected in the decisions that firms producing video game consoles and game software make regarding (backward) compatibility of their products [70]. Once the dominant player in this field, Sega effectively set its $s$ to zero when it decided not to make its new game console, the 32-bit Saturn, backward compatible to video games and complementary products that the hugely successful 16-bit Genesis had relied on. To challenge the subsequent dominant position of Sony's Playstation, Microsoft enfranchised Sega's portfolio of successful video game titles as Sega had by then ceased to produce consoles. Microsoft tried to tap into the network effects associated with Sega's portfolio of games and associated with the use of its operating system [70, p.13] when it launched its Xbox game console.

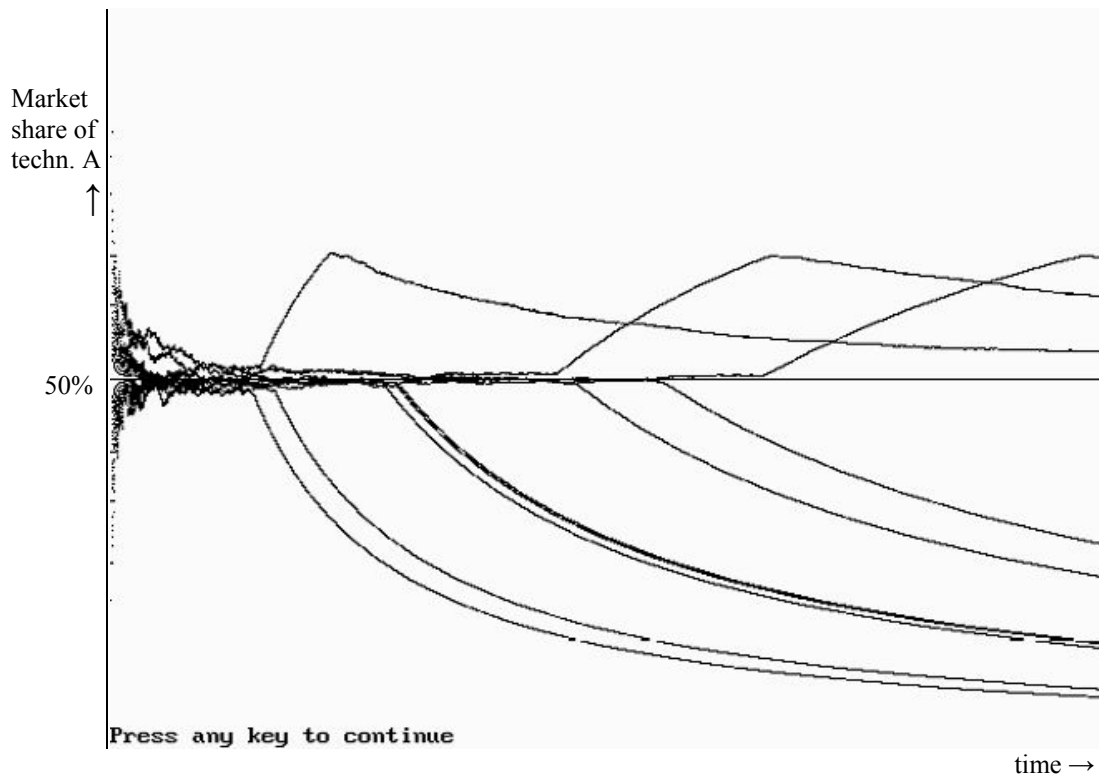

**Figure 4:** If S agents no longer profit from the network externalities for losing B technology ($s = 0$) lock-in to prevailing technology (A) is broken [31, p. 317).

A government (competition authority) or dominant player can, for example, demand from the Microsoft Company to publish the source code of its operating system, allowing others to design their product such that they will interface with it. A government may also actively provide information about the Linux operating systems. For router software network effects of the kind analyzed by Arthur may be weaker then for software that private individuals buy and use since the actors acquiring it are more knowledgeable and the software itself is compatible with more kinds of software at the interfaces. Due to the way in which Linux is developed, however, there is little incentive for producers to actively market Linux, and there is a low likelihood that consumers will actually be knowledgeable about this complex product. Government intervention, by providing information about open source software, to prevent undue lock-in may then be warranted as it increases social welfare [71]. A government may for similar reasons consider mandatory adoption by public agencies (ibid.). These scholars make their argument independent of the question whether or not Linux is technically superior as the jury on this matter is still out [see, e.g., 72]. These kinds of interventions may be sufficient for a competitive balance to remain or be restored.

The market share of a technology which has lost the competition can, paradoxically perhaps, not be improved in terms of current competitive circumstances. As Yahoo did, giving away its browser after it had lost the browser war with Microsoft's Explorer [73], one may have to destroy one's own network effects and related competences thus changing the nature of the competitive relations. Dissolution of the network effects may imply or result in a radical innovation, as the technology is freed from those contextual factors coupled with the development of the technology in the previous situation. Such a step down changes the selection environment for both technologies, possibly freeing up space for re-entry into the market by the firm backing up the previously locked-out technology.

In a corporate world, a shift of trajectory may require a different set of corporate alliances [74] or, in anticipation of a possible need to break-out, the development of a broader set of (technical) competences than is actually required in extant market conditions [75]. Since the technological restructuring is radical, it is based on a reconstruction and not on existing practices. One may have to abstract from the current situation and invoke a different knowledge base for making decisions. Firms that have distributed capabilities seem best equipped for this because the coordination problem is then considered as a cognitive puzzle (ibid.). Analyzing such strategic policy options for governments and firms in the light of the model of the lock-in of a technology developed enables us to understand which policy may actually hope to accomplish its objectives, and helps in devising additional, complementary instruments. It also allows one to understand counter-intuitive effects of strategic choices.

A techno-economic system can increasingly develop along a trajectory as two sub-systems interact at their respective interfaces. The system may become locked-into a suboptimum because the fitness landscape may be rugged [54, 55] as long as a third selection environment is relatively stable. The reaction-diffusion dynamics enables us to understand how a lock-in between a single technology and the market dynamics can be dissolved in a later stage. A co-evolution along a single trajectory can be "unlocked" when the diffusion mechanism of the market no longer co-evolves with the technical means of production. When another selection environment becomes relevant to a previously locked-in system, the new configuration may begin to tilt the system as soon as diffusion at the new interface becomes more important than, as suggested by Eq. 7, half of the rate along the trajectory of the system. Because an economic production system is attracted by market opportunities, one can expect a trajectory to be exploited to gain market share. The lock-in can thus be expected to erode as the diffusion rate for the new technology increases. A third selection environment, for example of political or strategic decision making, can become relevant, and reaction-diffusion dynamics may then open the lock-in.

## 5. Conclusion

Lock-in may result from the co-evolution of two selection environments or sub-systems, such as between the economic and the technological environment. Break-out from a lock-in is possible, but has not been modeled so far in the literature with a few exceptions [21]. This paper systematizes the discussion about the circumstances for break-out from a lock-in by arguing that when a third selection environment interacts with two locked-in ones, break-out or the return to a competitive balance is possible [58, 76]. Introduction of the reaction-diffusion mechanism into the model enables us to understand how such a process of break-out can be analyzed as an endogenous outcome of the dynamics of a system. It suggests conditions for the break-out of a lock-in or for return to a competitive balance between two technologies. The model developed allows one to understand better what options agents have already been experimented with actually do, and may suggest additional and possibly complementary ones, as lock-in, break-out or a competitive balance between technologies is sought.

**Acknowledgement**
We are grateful to two anonymous referees and the editor for comments and suggestions. The usual disclaimer holds, however.